\documentclass[twocolumn,prb]{revtex4}

\usepackage{graphicx}
\usepackage{dcolumn}
\usepackage{bm}

\begin{document}
\title{Chiral $d+is$ superconducting state in the two dimensional
$t$-$t^\prime$ Hubbard model}

\author{J. Mr\'az$^{1}$ and R. Hlubina$^{1,2}$}

\affiliation{
$^1$Department of Solid State Physics, Comenius University,
Mlynsk\'{a} Dolina F2, 842 48 Bratislava, Slovakia\\
$^2$International School for Advanced Studies (SISSA), 
Via Beirut 2-4, I-34014 Trieste, Italy}

\begin{abstract}
Applying the recently developed variational approach to Kohn-Luttinger
superconductivity to the $t$-$t^\prime$ Hubbard model in two
dimensions, we have found, for sizeable next-nearest neighbor hopping,
an electron density controlled quantum phase transition between a
$d$-wave superconducting state close to half filling and an $s$-wave
superconductor at lower electron density. The transition occurs via an
intermediate time reversal breaking $d+is$ superconducting phase,
which is characterized by nonvanishing chirality and density-current
correlation.  Our results suggest the possibility of a bulk time
reversal symmetry breaking state in overdoped cuprates.
\end{abstract}
\pacs{PACS}
\maketitle

As demonstrated by Kohn and Luttinger (KL), the generic ground state
of a clean degenerate electron system should be
superconducting.\cite{Kohn65} Direct application of the KL approach to
the $t$-$t^\prime$ Hubbard model in two dimensions has shown that, at
sufficiently weak coupling when no particle-hole instabilities
develop, the phase diagram of the model in the $t^\prime/t$ versus
electron density, $\rho$, plane, exhibits several superconducting
phases of different symmetry.\cite{Hlubina99} Similar conclusions have
been reached previously in mean field studies of the
$t$-$t^\prime$-$J$ model\cite{Koltenbah97} and more recently making
use of the renormalization group
method\cite{Shankar94,Zanchi97,Honerkamp01} and fluctuation exchange
approximation.\cite{Aoki03} Thus, by tuning one control parameter
($t^\prime/t$ or $\rho$), a quantum phase transition between two
different superconducting states can be realized.  The purpose of this
paper is to study such transitions within a simple but realistic
microscopic model.

As a particular example motivated by the physics of high temperature
superconductors, we have chosen to study the two-dimensional
$t$-$t^\prime$ Hubbard model with the bare single particle dispersion
$\varepsilon_{\bf k}=-2t(\cos k_xa+\cos k_ya)+4t^\prime\cos k_xa \cos
k_ya$ where $a$ is the in-plane lattice constant. This is the minimal
model including electron correlations and reproducing the shape of the
Fermi surface observed by angle resolved photoemission
spectroscopy\cite{Damascelli03} in the cuprates.  In a previous
study,\cite{Hlubina99} we have shown that close to half filling
$d$-wave is the dominant pairing symmetry in a broad range of
$t^\prime/t$ and that, under reducing the electron concentration,
different pairing symmetries start to dominate. Here we have chosen to
study the model for $t^\prime/t=0.45$, in which case a
density-controlled transition between the $d$-wave and $s$-wave
pairing states was predicted to occur in the vicinity of $\rho\approx
0.7$.\cite{Hlubina99} This choice of $t^\prime/t$ is motivated by the
observation\cite{Hlubina99} that at smaller ratios of $t^\prime/t$,
the transition either happens at small electron fillings, or is
located close to the Van Hove density in which case our method does
not apply. Moreover, according to a recent study,\cite{Pavarini01} Tl
and Hg based materials do indeed support sizeable next nearest
neighbor hopping.

We address this problem within the recently developed variational
approach to superconductivity.\cite{Mraz03} The main idea of
Ref.~\onlinecite{Mraz03} is as follows: Performing a canonical
transformation which eliminates the scattering processes to first
order in $U/t$ (where $U$ is the local Hubbard interaction), we have
constructed an effective Hamiltonian which leads to an interaction in
the Cooper channel of the form $V_{{\bf k}{\bf p}}=U+U^2 \chi_{1}({\bf
k}+{\bf p}, \varepsilon_{\bf p}-\varepsilon_{\bf k})$, where
$\chi_{1}({\bf q},\omega)$ is the real part of the particle-hole
susceptibility $\chi({\bf q},\omega)=L^{-1}\sum_{\bf K} (f_{\bf
K}-f_{{\bf K}+{\bf q}})/ (\varepsilon_{{\bf K}+{\bf
q}}-\varepsilon_{\bf K}-\omega-i0)$ and $L$ is the number of lattice
sites. Superconductivity is realized if a nontrivial order parameter
$\Delta_{\bf k}$ which solves the gap equation
\begin{equation}
\Delta_{\bf k}=-{1\over L}\sum_{\bf p}V_{{\bf k}{\bf p}}
\Delta_{\bf p}{\tanh(E_{\bf p}/2T)\over 2E_{\bf p}}
\label{eq:gap}
\end{equation}
can be found. In Eq.~(\ref{eq:gap}) we introduced the quasiparticle
enegy $E_{\bf p}=(\xi_{\bf p}^2+|\Delta_{\bf p}|^2)^{1/2}$, where
$\xi_{\bf p}=\varepsilon_{\bf p}-\mu$ and $\mu$ is the chemical
potential.

All calculations reported in this paper were performed for moderate
interaction strength $U/t=7$. Strictly speaking, this is outside the
region of validity of our weak-coupling approach. The need for such
values of $U$ is due to the finite system sizes, typically up to
$512\times 512$, which we can study within a reasonable time. At
smaller values of $U$, the finite size effects become appreciable.
However, we do not believe the large values of $U$ are a serious
drawback of our calculations. In fact, in the $d$-wave state the value
of $U$ is just a multiplicative factor entering $V_{{\bf k}{\bf p}}$
(since the $d$-wave state is not affected at all by the constant term
$U$ in $V_{{\bf k}{\bf p}}$). On the other hand, although both terms
in $V_{{\bf k}{\bf p}}$ do affect the $s$-wave pairing state, we have
checked a posteriori that our solutions in the $s$-wave channel
exhibit very small on-site pairing amplitudes, thus providing a
nontrivial self-consistency check of the calculation.

\begin{figure}[t]
\centerline{\includegraphics[width=6.3cm,angle=0]{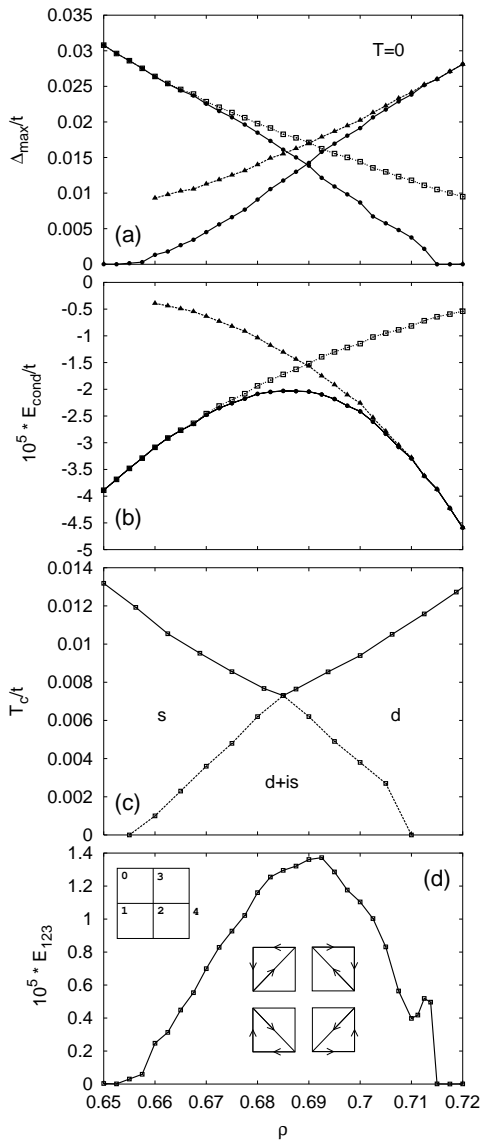}}
\caption{\label{fig:S1} (a) Maximal value of $|\Delta_{\bf k}|$ at the
Fermi surface. Triangles: pure $d$-wave state. Squares: pure $s$-wave
state. Circles: the same for $|d_{\bf k}|$ and $|s_{\bf k}|$ in the
$d+is$ state.  (b) Condensation energy per lattice site (in units of
$t$) in the pure $d$-wave (triangles), pure $s$-wave (squares), and in
the $d+is$ state (circles).  (c) Phase diagram of the $t$-$t^\prime$
Hubbard model.  (d) Spin chirality $E_{123}$.  The upper inset defines
the lattice points 0,1,2,3, and 4.  The lower inset shows the pattern
of current chirality on the plaquette 0123.  All data were calculated
as functions of the electron density $\rho$ for $t^\prime/t=0.45$ and
$U/t=7$ on special lattices \cite{Mraz03} with $L=512\times 512$ (a,d)
and $L=256\times 256$ (c). (b) was calculated by extrapolating the
results in (a) to $4096\times 4096$.  }
\end{figure}

At $\rho=0.685$ we have solved the gap equation at $T=0$ by an
iteration method. The calculation was performed many times, starting
from a random complex vector $\Delta_{\bf k}$.  We have found three
types of solutions: pure $d$-wave, pure $s$-wave, and $d+is$-type
solutions which can be written as $\Delta_{\bf k}=d_{\bf k}+is_{\bf
k}$ with real $d_{\bf k}$ and $s_{\bf k}$.  As a next step, we have
required that the self-consistent solution is one of the above three
types (which speeded up the calculation considerably) and in this way
we calculated the data presented in Fig.~\ref{fig:S1}a. Note that pure
$s$-wave and $d$-wave solutions are stable for all studied densities,
whereas the $d+is$-type solution can be stabilized only in the
interval $0.655<\rho<0.715$.

We have calculated the BCS condensation energy $E_{\rm cond}=
-\sum_{\bf k}{(E_{\bf k}-|\xi_{\bf k}|)^2/ 2 E_{\bf k}}$ as a function
of $\rho$ for all three types of solutions. The results are plotted in
Fig.~\ref{fig:S1}b. We find that the $d+is$ state is the true ground
state in the whole region of its stability.  Thus, at weak coupling,
the quantum phase transition between the $s$-wave state at low
electron density and the high-density $d$-wave state occurs via a
mixed symmetry state $d+is$, in agreement with a model calculation for
an isotropic Fermi surface and separable
interactions.\cite{Musaelian96} Within our mean field like approach,
both $s$/$d+is$ and $d$/$d+is$ transitions are of second order.  Field
theoretic studies suggest that for the $d$/$d+is$ transition this is
valid also beyond mean field.\cite{Vojta00,Khveshchenko01}

\begin{figure}[t]
\centerline{\includegraphics[width=7.0cm,angle=0]{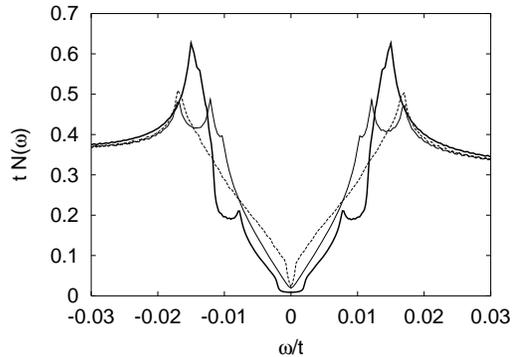}}
\caption{\label{fig:DOS} Density of states per lattice site (in units
of $t$) for $\rho=0.69$ in the pure $d$-wave, pure $s$-wave, and in
the $d+is$ state (dashed, thin solid, and thick solid lines,
respectively). Calculated following Ref.~\onlinecite{Mraz03} for the
same parameters as Fig.~\ref{fig:S1}. The finite weight at low
energies is due to the finite width of the $\delta$ functions,
$\gamma/t=2\times 10^{-4}$. }
\end{figure}

\begin{figure}[t]
\centerline{\includegraphics[width=8.5cm,angle=0]{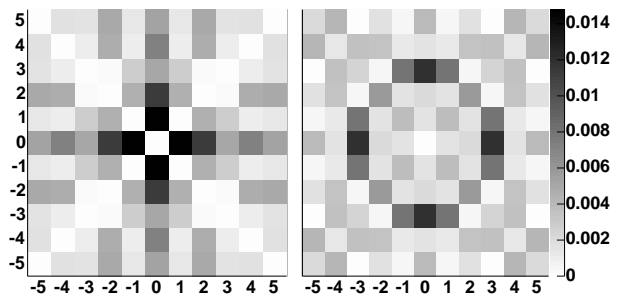}}
\caption{\label{fig:pairing} Magnitudes of the real (left panel) and
imaginary (right panel) parts of the pairing function, $|D_{ij}|$ and
$|S_{ij}|$, in the chiral $d+is$ state. The middle of the square
corresponds to $i=j$.  Calculated for $\rho=0.69$ and the same
parameters as in Fig.~\ref{fig:S1}a.}
\end{figure}

It is worth pointing out that the complex admixture increases the
condensation energy substantially, up to 30 per cent close to the
crossing point of pure $s$-wave and $d$-wave states. Since the
condensation energy is determined by the typical value of
$|\Delta_{\bf k}|$ close to the Fermi energy and, according to
Fig.~\ref{fig:S1}a, the maximal value of $|\Delta_{\bf k}|$ is
approximately the same in the mixed and pure states close to the
crossing point, the density of states in the $d+is$ state must be
substantially suppressed in the subgap region. This is indeed found in
an explicit calculation, see Fig.~\ref{fig:DOS}. Note that a small
true gap is opened in the $d+is$ state, whereas both the $d$-wave and
the $s$-wave states are gapless.\cite{s_wave}

In order to get further insight into the nature of the $d+is$ state,
let us consider the singlet pairing function
$F_{ij}=\sum_\sigma\sigma\langle c_{i-\sigma}c_{j\sigma}\rangle$.  In
the $d+is$ state we can write $F_{ij}=D_{ij}+iS_{ij}$ where
$D_{ij}=L^{-1}\sum_{\bf k}(d_{\bf k}/E_{\bf k}) \cos{\bf k}\cdot{\bf
R}_{ij}$, $S_{ij}= L^{-1}\sum_{\bf k}(s_{\bf k}/ E_{\bf k}) \cos{\bf
k}\cdot{\bf R}_{ij}$, and ${\bf R}_{ij}$ is the vector connecting
lattice sites $i$ and $j$.  Note that $D_{ij}$ and $S_{ij}$ are real
and even in $i$,$j$. In Fig.~\ref{fig:pairing} we plot $|D_{ij}|$ and
$|S_{ij}|$ in the chiral state. We find that $|D_{ij}|$ is peaked for
$j-i=(1,0)$, $(2,0)$, and symmetry related points, whereas
$|S_{ij}|$ peaks for $j-i=(3,0)$, $(3,1)$, $(2,2)$, and symmetry
related points. This means that the $d$-wave and $s$-wave states do
not compete for the same Cooper pairs, allowing for a coexistence of
both states in the $d+is$ state.

\begin{figure}[t]
\centerline{\includegraphics[width=7.0cm,angle=0]{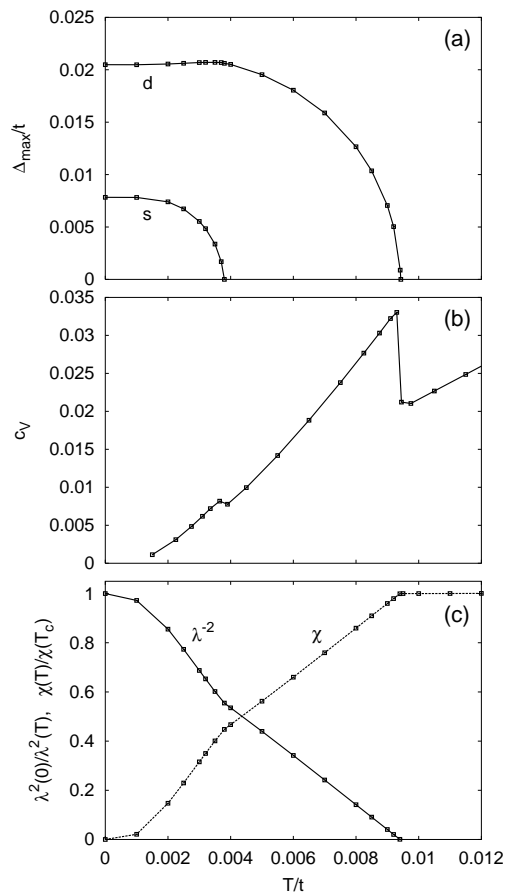}}
\caption{\label{fig:Delta_temp} (a) Maximal amplitudes of the real and
imaginary parts of $\Delta_{\bf k}$ at the Fermi surface in the $d+is$
state, (b) specific heat $c_V$ per lattice site, and (c) normalized
penetration depth and spin susceptibility as functions of temperature.
Calculated for $\rho=0.7$ (and the same parameters as in
Fig.~\ref{fig:S1}) on a special lattice\cite{Mraz03} with $L=256\times
256$.  (b) and (c) were calculated by extrapolating the data from (a)
to a lattice $1024\times 1024$.}
\end{figure}

The temperature dependence of the $d$-wave and $s$-wave parts of the
gap in the $d+is$ state is shown for $\rho=0.7$ in
Fig.~\ref{fig:Delta_temp}a.  Two critical temperatures $T_c^d$ and
$T_c^{d+is}$ can be defined, the mixed symmetry state being realized
at $T<T_c^{d+is}$, and the pure $d$-wave state occuring at
$T_c^{d+is}<T<T_c^d$. Collecting the data for different $\rho=$~const
cuts together we obtain the phase diagram Fig.~\ref{fig:S1}c.  Note
that the region of stability of the $d+is$ state shrinks with
increasing temperature, due to the lower entropy in the mixed state
than in the pure states. 

The complex order parameter in the $d+is$ state implies that time
reversal symmetry is broken.  Moreover, the point group symmetry is
reduced from $C_{4v}$ down to $C_{2v}$, i.e. the 90$^\circ$ rotations
and reflections across the diagonals, $m^\prime$, do not belong any
more to the symmetry operations of the $d+is$ state.  A physical order
parameter which distinguishes the two possible time-reversal related
states $d+is$ and $d-is$ can be constructed as an odd function of the
local spin operator ${\bf S}_i=2^{-1}{\hat \sigma}_{\alpha\beta}
c^\dagger_{i\alpha}c_{i\beta}$ (${\hat \sigma}$ is the vector of Pauli
matrices) or of the charge current
$j_{ij}=-it_{ij}\sum_\sigma(c^\dagger_{i\sigma}c_{j\sigma}
-c^\dagger_{j\sigma}c_{i\sigma})$ defined on the bond
$ij$.\cite{hopping} However, because of spin rotation invariance of the
$d+is$ state, we have $\langle {\bf S}_i\rangle=0$, while the unbroken
$m$ reflection symmetry, i.e. $x\rightarrow -x$, forces $\langle
j_{ij}\rangle=0$.  Therefore more complicated functions of ${\bf S}_i$
and $j_{ij}$ have to be considered.

Let us start by taking the simplest order parameter made up solely of
the spin operators. This must necessarily be a third power of ${\bf
S}_i$. The spin chirality\cite{Wen89} $E_{ijk}=\langle{\bf
S}_i\cdot\left({\bf S}_j\times{\bf S}_k\right)\rangle$, where $i$,
$j$, and $k$ are lattice sites, is a natural candidate for such an
order parameter, since it is spin rotation invariant.  Very recently,
the $d+id_{xy}$ RVB spin liquid has been characterized\cite{Sorella03}
by $E_{ijk}$ and in what follows we show that this order parameter is
useful in our case as well, if the lattice points $i$, $j$, and $k$
are chosen in such a way that a nonzero value of $E_{ijk}$ is not
prohibited by the remaining $C_{2v}$ symmetry.

In a singlet superconducting state, the chiral order parameter
can be evaluated as
\begin{eqnarray}
E_{ijk}=
-{3\over 4}{\rm Im}\left[G_{ij}F_{jk}F_{ki}^\ast+
G_{jk}F_{ki}F_{ij}^\ast+G_{ki}F_{ij}F_{jk}^\ast\right],
\label{eq:chirality}
\end{eqnarray}
where $G_{ij}=\langle
c^\dagger_{i\sigma}c_{j\sigma}\rangle=L^{-1}\sum_{\bf k}f_{\bf
k}\cos{\bf k}\cdot{\bf R}_{ij}$ is independent of $\sigma$ and an even
function of $i$ and $j$.  For convenience let us introduce the
following abbreviated notation for the neighborhood of the lattice
point ${\bf R}$ (see inset to Fig.~\ref{fig:S1}d): $0={\bf R}-{\bf
x}+{\bf y}$, $1={\bf R}-{\bf x}$, $2={\bf R}$, $3={\bf R}+{\bf y}$,
and $4={\bf R}+{\bf x}$, where ${\bf x}$ and ${\bf y}$ are unit
lattice vectors in the $x$ and $y$ direction, repectively.  Consider
first the triangle 123.  Since $G_{12}=G_{23}$, $S_{12}=S_{23}$,
$D_{12}=-D_{23}$, and $D_{13}=0$, making use of
Eq.~(\ref{eq:chirality}) one finds readily that
$E_{123}=(3/2)D_{12}(S_{12}G_{13}-S_{13}G_{12})$. In
Fig.~\ref{fig:S1}d we plot $E_{123}$ as a function of doping.

Let us point out that the chiral order parameter provides also a
direct measure of broken spatial symmetries.  In fact, the
nonvanishing value of $E_{123}$ is a direct consequence of broken
$m^\prime$ symmetry, since in a $m^\prime$ symmetric system
$E_{123}=E_{321}=-E_{123}$. Moreover, a clockwise rotation of the
lattice around the point 2 leads to $E_{123}\rightarrow E_{324}$, and
a straightforward calculation shows $E_{324}=-E_{123}$. On the other
hand, under the $m$ reflection we have $E_{123}\rightarrow
E_{423}=E_{123}$, consistent with the invariance of the $d+is$ state
under the $m$ reflection.

\begin{figure}[t]
\centerline{\includegraphics[width=7.5cm,angle=0]{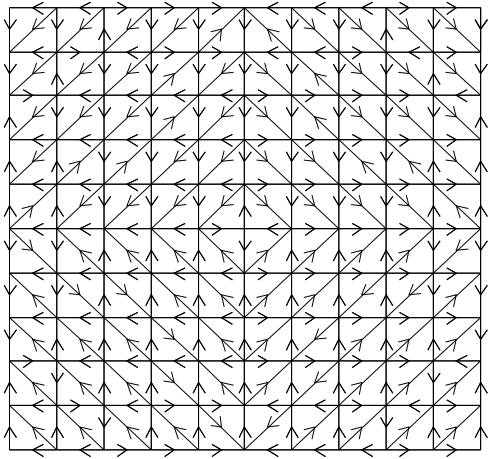}}
\caption{\label{fig:B} Density-current correlation function
$B_{ijk}$. Lattice point $i$ is in the center. The arrow points from
$j$ to $k$ for $B_{ijk}>0$ and from $k$ to $j$ otherwise. For clarity,
only half of the diagonal currents is shown. Calculated from the
pairing function in Fig.~\ref{fig:pairing}.}
\end{figure}

Let us turn to order parameters based on current operators. The
simplest order parameter of this type is\cite{Anderson61}
$$
B_{ijk}={1\over 2}\langle n_ij_{jk}+j_{jk}n_i\rangle
=t_{jk}{\rm Im}(F^\ast_{ij}F_{ik}),
$$ and the equality applies to a general singlet superconductor. For a
$d+is$ superconductor, this specializes to
$B_{ijk}=t_{jk}\left(D_{ij}S_{ik}-S_{ij}D_{ik}\right)$.  An example of
$B_{ijk}$ is shown in Fig.~\ref{fig:B}.  The nonvanishing
density-current correlation $B_{ijk}$ suggests that potential
scattering on imperfections such as impurities or surfaces will, in
general, cause finite currents and magnetic fields in the vicinity of
these defects. For scattering on nonmagnetic impurities, such currents
have been explicitly calculated only recently (for a $d+id_{xy}$
superconductor).\cite{Salkola98} On the other hand, nonvanishing
surface currents have been predicted already by Anderson and
Morel.\cite{Anderson61} In our case the surface currents should be
most pronounced for $[110]$-like surfaces, while for $[100]$-like
surfaces the unbroken $m$ symmetry forbids the formation of such
currents. This is fully consistent with the predictions of
Ginzburg-Landau theory.\cite{Sigrist98}

Next let us consider, in analogy to spin chirality, the third power of
local current operators, $P_{ijk}=\langle
j_{ij}j_{jk}j_{ki}-j_{ik}j_{kj}j_{ji}\rangle$. Note that, being an
expectation value of a Hermitian operator
$j_{ij}j_{jk}j_{ki}+(j_{ij}j_{jk}j_{ki})^\dagger$, the order parameter
$P_{ijk}$ is a real quantity. A nonzero value of $P_{ijk}$ means that
the product of three current operators along a closed loop is
different for current operators ordered clockwise and anticlockwise
around the loop. Therefore we call $P_{ijk}$ the current chirality.  A
lengthy but straightforward calculation shows that, in a singlet
superconductor, $P_{ijk}=-(32/3)t_{ij}t_{jk}t_{ki}E_{ijk}$.  Thus, up
to a trivial multiplicative constant, the spin and current chiralities
are equal. Making use of the symmetry properties of $E_{123}$ we find
that the four possible current chiralities which can be defined on an
elementary plaquette 0123 of the square lattice satisfy
$P_{103}=P_{012}=-P_{032}=-P_{123}$.  The resulting current chirality
pattern is shown in the inset of Fig.~\ref{fig:S1}d.  Note the
similarity of Fig.~\ref{fig:S1}d with the distribution of currents in
a chiral state of a three-band model introduced by Varma as a
description of the pseudogap phase of the cuprates.\cite{Varma97} This
similarity is consistent with the recent symmetry classification of
superconducting states coexisting with a chiral translationally
invariant pseudogap phase of nonsuperconducting origin.\cite{Kaur03}
The crucial difference of our calculation with respect to
Ref.~\onlinecite{Varma97} is that in our case no net currents are
flowing, $\langle j_{ij}\rangle=0$. This makes it possible that a
translationally symmetric chiral state is realized also in a one band
model.

Our weak coupling approach, which should be reliable sufficiently away
from half filling, therefore predicts a bulk chiral $d+is$ state in
overdoped cuprates with sizeable next-nearest neighbor hopping.  Next
we address the question about experimental observability of such a
state. Upon cooling, two phase transitions should be visible.
Figure~\ref{fig:Delta_temp}b shows that this is indeed the case but
the anomaly of the specific heat at $T_c^{d+is}$ is small, since the
quasiparticles are already frozen out to a large extent at
$T_c^{d+is}$.  This is expected to be a generic result which holds
except for the case when $T_c^d\sim T_c^{d+is}$.  On the other hand,
within BCS theory the in-plane penetration depth is given by
\begin{eqnarray*}
\lambda^{-2}&=&
{e^2\over \varepsilon_0da^2\hbar^2c^2}
{1\over L}\sum_{\bf k} 
\left({\partial\varepsilon_{\bf k}\over\partial {\bf k}}\right)^2
\left[{\partial f(E_{\bf k})\over\partial E_{\bf k}}
-{\partial f(\xi_{\bf k})\over\partial \xi_{\bf k}}\right],
\end{eqnarray*}
where $d$ is the interlayer distance.  In Fig.~\ref{fig:Delta_temp}c
we show explicitly that $T_c^{d+is}$ should be observable in the
temperature dependence of $\lambda$, and also of the spin
susceptibility $\chi$. Note also the activated behavior at low
temperatures which is consistent with the full gap.

Direct tests of time-reversal symmetry breaking are more involved,
since in a perfect sample $\langle j_{ij}\rangle=0$ and therefore no
spontaneous magnetic fields develop. Moreover, since the photoemission
experiments test only the diagonal part of the Nambu-Gorkov electron
Green's function, the broken parity in our $d+is$ state can not be
directly measured by circular dichroism\cite{Varma97} of photoemission
spectra.

Recently a spontaneous splitting of zero bias conductance peaks in a
tunneling experiment on overdoped cuprates has been interpreted in
terms of a bulk time reversal symmetry breaking superconducting
state.\cite{Dagan01} However, this experiment measures the properties
of a [110] oriented surface of a $d$-wave superconductor and in this
geometry, time reversal symmetry may be broken in the vicinity of the
surface even in absence of such symmetry breaking in the
bulk.\cite{Sigrist98} Therefore phase sensitive experimental tests of
the putative bulk chiral $d+is$ state at large doping should avoid
geometries with in-plane edges. We propose to measure the
current-phase relation $I(\varphi)$ of $c$-axis Josephson junctions
between a conventional superconductor and overdoped cuprates. Assuming
that the Josephson coupling is due to a finite $s$ component of the
cuprates, from the relative sign of the first and second harmonics of
$I(\varphi)$ one can determine the relative phase of the $d$ and $s$
components of the order parameter.\cite{Hlubina02}

In conclusion, applying the recently developed variational approach to
KL superconductors,\cite{Mraz03} we have shown that for sufficiently
large next-nearest-neighbor hopping $t^\prime/t$, the Hubbard model
predicts a change of pairing symmetry from $d$-wave to $s$-wave with
overdoping, in agreement with a previous calculation.\cite{Hlubina99}
The quantum phase transition between the two superconducting states
occurs via an intermediate chiral $d+is$ phase. We have shown that the
density-current correlation and the spin and current chiralities serve
as local measures of the broken time reversal and parity symmetry in
the $d+is$ state, and we calculated their spatial distribution.  Our
results suggest the possibility of a bulk time reversal symmetry
breaking state in overdoped cuprates with sizeable next-nearest
neighbor hopping.  We have calculated thermodynamic quantities which
might be useful for experimental tests of the putative $d+is$ pairing
state.

We thank M. Mo\v{s}ko for suggestions on how to improve the
convergence of the self-consistent calculation.  RH thanks S.~Sorella
for discussions and the Italian Ministry for Education, Universities,
and Research for financial support.  This work was supported in part
by the Slovak Scientific Grant Agency under Grant No.~VEGA-1/9177/02
and by the Slovak Science and Technology Assistance Agency under the
contract No.~APVT-20-021602.


\begin{thebibliography}{99}
\bibitem{s_wave}For $\rho\approx 0.7$, the gap in the $s$-wave and
$d$-wave states varies along the Fermi line approximately as $\cos
8\varphi$ and $\sin^2 2\varphi \cos 2\varphi$, respectively, where
$\varphi$ is the polar angle.
\bibitem{hopping}Here we have assumed real hopping matrix elements
$t_{ij}=t_{ji}$. In particular, we have $t_{ij}=t$ for nearest
neighbor and $t_{ij}=-t^\prime$ for next nearest neighbor sites $i$
and $j$.
\end{thebibliography}
\end{document}